\documentclass[12pt]{article}
\usepackage{epsfig}
\begin{document}
\thispagestyle{empty}
\setcounter{page}{0}
\renewcommand{\theequation}{\thesection.\arabic{equation}}
\begin{quote}
{\hfill{\tt hep-th/0406162}}

{\hfill{ULB-TH/04-19}}
\vskip 1.5cm

\begin{center} \Large { \bf   Broken symmetry and Yang-Mills theory}\footnote{ Contribution to ``Fifty
years of Yang Mills theory'', editor G. 't Hooft, World Scientific.}

\vskip.5cm
\small

\hfill ``{\it Only by their breaking could the divine configurations be perfected''}

\footnotesize
\hfill  Kabbalistic text; Ta'alumoth Chokhmah (The Channels of Wisdom) 1629

\hfill Joseph Salomon del Medigo of Crete
\vskip2cm\normalsize
 Fran\c cois Englert
\vskip .5cm \footnotesize{\em Service de Physique Th\'eorique\\ Universit\'e Libre de   Bruxelles,
  Campus Plaine, C.P.225\\ and\\The International
Solvay Institutes, Campus Plaine C.P. 231\\ Boulevard du Triomphe, B-1050 Bruxelles, Belgium}\\
 {\tt
fenglert@ulb.ac.be}
\end{center}

\vskip 3cm
\small
\centerline{\bf Abstract}

 From its inception in statistical physics to its role in the construction and in the
development of the asymmetric Yang-Mills phase in quantum field theory, the notion of spontaneous broken
symmetry permeates contemporary physics.  This  is reviewed with particular emphasis on the
conceptual issues.
\end{quote}
\newpage
\setcounter{equation}{0}
\addtocounter{footnote}{-1}
\small
\baselineskip 18pt
\section{Introduction}
 Physics, as we know it, attempts to interpret the  diverse natural phenomena as particular
manifestations of general laws. The impressive success of this enterprise in the first half of the
twentieth century made it conceivable that  all phenomena at the atomic level and at larger distance
scales be governed solely by the known laws of classical general relativity and quantum
electrodynamics.

The vision of a world ruled by general testable  laws is relatively recent. Basically it was initiated 
by the Galilean inertial principle. The subsequent rapid development of large-scale physics is
certainly tributary to the fact that gravitational and electromagnetic forces are long-range and 
hence can be perceived directly without the mediation of highly sophisticated technical devices.  

The discovery of subatomic structures and of the concomitant weak  and strong  short-range forces  
raised  the question of how  to cope with  short-range forces in relativistic quantum field theory. The
Fermi theory of weak   interactions, formulated in terms of a four Fermi point-like current-current
interaction, was well-defined in lowest order perturbation theory and successfully confronted many
experimental data. However, it is clearly inconsistent in higher orders because of uncontrollable
divergent quantum fluctuations. In order words, in contradistinction to quantum electrodynamics,
 the Fermi theory is not renormalizable. This
difficulty could not be solved by smoothing the point-like interaction by a massive, and therefore
short-range, charged vector particle exchange (the so-called $W^+$ and $W^-$ bosons): theories
with  massive  charged vector bosons are  not renormalizable either. In the early nineteen sixties,
there seemed to be insuperable obstacles to  formulating  a  consistent theory with short-range
forces mediated by massive vectors. 

It is the notion of spontaneous broken symmetry as adapted to  gauge theory that provided the clue
to the solution. 

The notion of spontaneous broken symmetry (SBS) finds its origin in the statistical physics of  phase
transitions \cite {landau}.  There, the low temperature ordered  phase of the system can be
asymmetric with respect to the symmetry principles that govern its dynamics. This is not
surprising since more often than not  energetic considerations dictate that the ground state or low
lying excited states of a many body system become ordered. A collective variable such as
magnetization picks up  expectation value, which define an order parameter   that  otherwise would
vanish by virtue of the dynamical symmetry (isotropy in the aforementioned example).  More
surprising was the discovery by Nambu that  the vacuum and the low energy excitations of a
relativistic field theory may bare the mark of SBS \cite{nambu2,nambujl}. Broken chiral symmetry
due to a spontaneous generation of hadron mass induces  massless pseudoscalar modes (identified
with a massless limit of  pion fields), which at infinite wavelength generate  rotation of the chiral
phase.  In absence of gauge field, such massless Nambu-Goldstone (NG) modes and  the concomitant
vacuum degeneracy in the coset $\cal G/\cal H$, where  $\cal G$ is the symmetry group and
$\cal H $ the unbroken subgroup,  are  general features of spontaneous broken symmetry of a
continuous group.  The occurrence of SBS,  of either a continuous or a discrete group, is also marked
by  fluctuations of the order parameter described by generically massive  scalar
excitations. Introducing gauge fields renders local in space-time the otherwise global dynamical
symmetry
$\cal G $  and leads to dramatic  effects.   While the massive   scalar excitations  survive,  the
massless NG bosons disappear as such but provide a longitudinal polarization for the gauge bosons
living in the coset, which become massive.  The essential degeneracy of the   vacuum 
is  removed and local gauge invariance is preserved despite the gauge boson masses.  Thus, the
apparent global  broken symmetry from  $\cal G$ to $\cal H $ is now hiding a true unbroken local
gauge symmetry.

This way of obtaining massive vectors and hence short-range forces out of a fundamental massless
Yang-Mills Lagrangian was proposed in 1964 independently by Brout and Englert in quantum field
theoretic terms~\cite{eb} and by Higgs   in  the equations of motion formulation~\cite {higgs2}. The
preserved gauge invariance was the  cornerstone, as in quantum electrodynamics although in a much
more sophisticated way,  of the proof by 't Hooft and Veltman that the mechanism of Brout, Englert
and Higgs (BEH) yields a renormalizable theory
\cite{renorm}. The renormalizability made entirely consistent the electroweak theory
\cite{gws}, proposed by Weinberg in 1967,  related to a group
theoretical model of Glashow and to the dynamics of  the BEH mechanism. 

I shall review the basic concepts leading to the construction of  gauge vector boson
masses, discuss  further developments  and their role  in contemporary
physics.

\setcounter{equation}{0}
\section{Spontaneous  broken global symmetry}
\subsection{Broken  symmetry in statistical physics}

Consider a condensed matter system, whose dynamics is invariant under a  continuous  symmetry.
As the temperature  is lowered below a critical one, the  symmetry  may be reduced by the
appearance of an ordered phase. The breakdown of the original symmetry is always  a discontinuous
event at the phase transition point but the order  parameters may set in continuously as a function
of temperature. In the latter case the phase transition is second order.  Symmetry breaking  by a
second order phase transition occurs in particular in ferromagnetism, superfluidity and
superconductivity. I first discuss in detail the ferromagnetic phase transition which illustrates
three general features of global SBS: ground state degeneracy, the appearance of a ``massless mode''
when the dynamics is invariant under a  continuous  symmetry,  and the occurrence  of a ``massive
mode''.

\vskip 1cm
\hskip -.3cm
\epsfbox{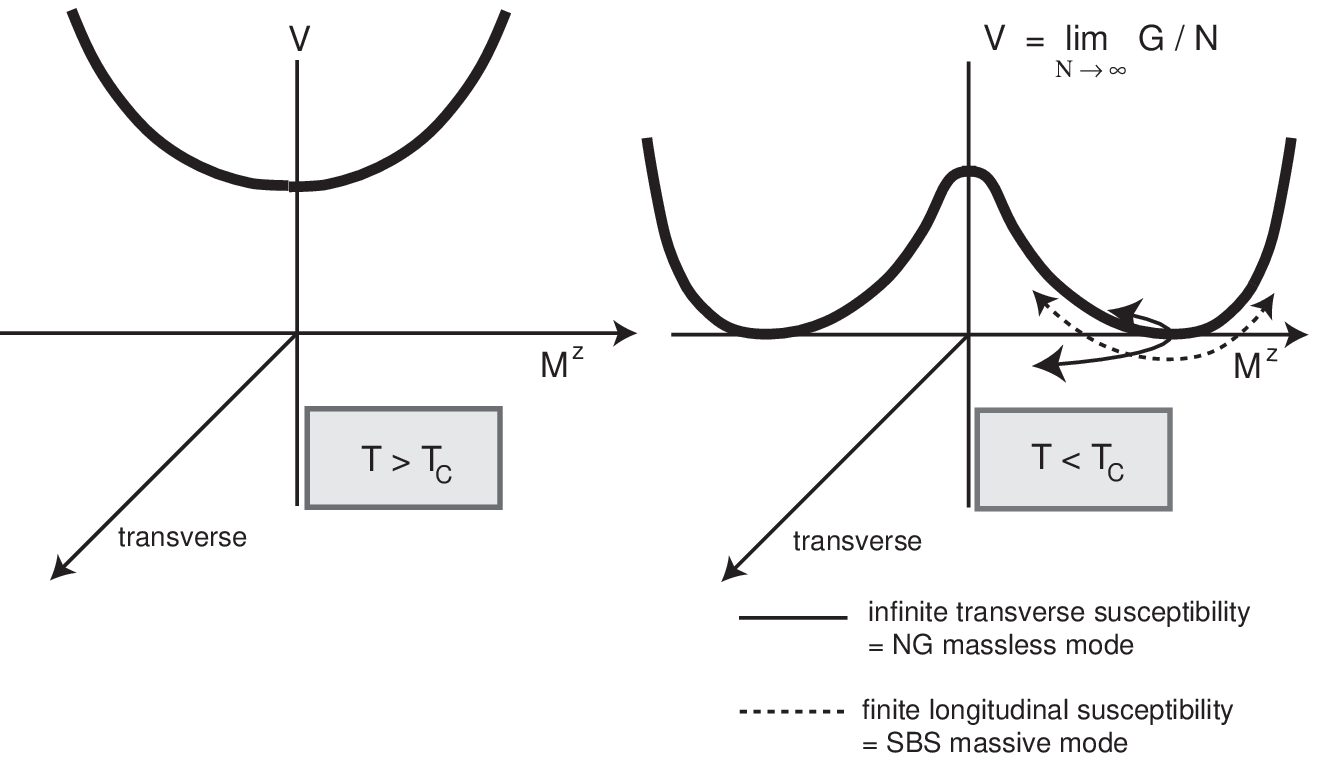}
\begin{quote}
\begin{center}
\baselineskip 12pt {\footnotesize Fig.1.  Effective potential of a Heisenberg ferromagnet.}
\end{center}
\end{quote}
\vskip .3cm

 In absence of external magnetic fields and of surface effects, a ferromagnetic substance below the
Curie point displays a global orientation of the magnetization, while the dynamics of the system is
clearly rotation invariant. This is SBS. The features of SBS are neatly illustrated by taking the
Heisenberg model with spin
$1/2$ defined by the Hamiltonian
\begin{equation}
\label{heisenberg} H= H_0 - \sum_{i=1}^N S_i^z h^z\qquad\qquad H_0=-2\sum_{i\neq j}v_{ij}{\vec
S}_{i} . {\vec S}_{j}\, ,
\end{equation} in the limit   $N\to\infty$.  Here $v_{ij}$ is the exchange potential between the spins
$\vec{S}_{i}\, ,\vec{S}_{j}$ located at the lattice sites
$i$ and $j$. $\vec S_i=\vec\sigma_i/2$ where the components of $\vec\sigma_i$ are the Pauli
matrices
$\{\sigma^x_i\, ,\sigma^y_i\, ,\sigma^z_i\} \, ([\vec\sigma_i,\vec\sigma_j] =0\,  \hbox{for}\, i\neq j)
$ and
$\vec h = h^z\vec1_z$ an applied magnetic field in the
$z$-direction.    Define the average magnetization   $\vec M = M^z\vec1_z$, $
M^z=\lim_{N\to\infty}(1/N)\langle
\sum_i S_i^z\rangle_{T, h^z}$. The effective potential $V$  is the Gibbs potential per spin
$G/N= (E-TS)/N+\vec M.\vec h $.  It is given by  $V=-kT\lim_{N\to\infty}(1/N)\ln Z_N+\vec M.\vec h $ ,
where $ Z_N={\rm Tr} \exp (-H/kT)$.  Its behavior is characteristic  of second  order phase transitions
with spontaneous broken symmetry~\cite{landau}. Above the Curie point, V has a single minimum at
$\vec M=0$. This minimum flattens at at $T=T_c$ and two symmetric minima appear for
$T<T_c$ in the $VM^z$-plane. This would be the whole story for a system with discrete symmetry,
such as the Ising model obtained from the Hamiltonian Eq.(\ref{heisenberg}) by retaining only the
$z$-component of the spin. The discrete $Z_2$ symmetry of the action would be 
spontaneously broken  below the Curie point when, as $h^z\to 0$, the system ends in one of the
  equivalent minima  in the $VM^z$-plane exhibited in Fig 1. 

But, when
$\vec h=0$, the  Hamiltonian $H_0$ is invariant under the full rotation group
$SO(3)$. This continuous  symmetry  implies that the thermodynamics of the ferromagnetic
phase does not depend on the orientation of the magnetization. The  effective
potential $V(T,M)$    only        depends  on the norm
$M$ of the magnetization vector $\vec M$.  Hence the equivalent minima are not only doubly
degenerate but span the full 2-sphere, that is the coset space of
$SO(3)/U(1)$. By selecting an  orientation at a given minimum, the system in the ferromagnetic phase
spontaneously breaks the
$SO(3)$ symmetry down to $U(1)$.

Consider now the system at $T=0$. 
The magnetization vector $\vec M$ has a quantum mechanical
interpretation.
 As $h^z\to 0$ the magnetization of the  ground state points in the
$z$-direction. It is  the symmetric ``all spin up'' state
$\vert 0\rangle =\vert +++\cdots
\rangle$ where the normalized spin states of the individual spins $\vert +\rangle$ are  quantized   in
the
$z$-direction. One easily verifies that $\vert 0\rangle$ is an eigenstate of $H$ and one has
\begin{equation}
\vec S \stackrel{\rm def}{=}  \sum_i \vec S_i\quad\quad  \langle 0 \vert S^x\vert 0\rangle= \langle
0
\vert S^y\vert 0\rangle=0\quad\quad \langle 0 \vert S^z\vert 0\rangle = NM=\frac{N}{2}\, . 
\end{equation} The operators $S^\alpha \, (\alpha = x,y,z)$ obey $[ S^\alpha, S^\beta] =i
\epsilon^{\alpha\beta\gamma} S^\gamma $ and are generators of the rotation group. One may 
construct the rotated ground states  from them. The   state $\vert\theta\rangle$ obtained from $\vert
0\rangle$  by rotating an angle $\theta$  about the $x$-axis is
\begin{equation}
\vert \theta\rangle = e^{iS^x\theta} \vert 0\rangle\, .
\end{equation} The states  $\vert
\theta\rangle$ and $\vert 0\rangle$ are degenerate since $[{\cal H}, S^x] =0  $ and one gets from the
commutation relations
\begin{equation}
\langle\theta\vert S^x\vert\theta\rangle=0 \quad\quad\langle\theta\vert
S^y\vert\theta\rangle=NM
\sin\theta\quad\quad\langle\theta\vert S^z\vert\theta\rangle=NM\cos\theta\, .
\end{equation} In this way, the classical notion of the ``arrow'' in $\vec M$ is given by the expectation
value of the operator $\vec S$ in the different rotated ground states.

Consider now the two distinct ground states, $\vert 0\rangle$ and $\vert
\theta\rangle$. One has
\begin{eqnarray}
\label{orthogonal}
\langle 0\vert\theta\rangle &=&\langle 0\vert e^{iS^x\theta} \vert 0\rangle = \langle 0\vert
\prod_{i=1}^N   e^{i(\sigma_i^x /2)\theta} \vert 0\rangle\nonumber\\ &=&\prod_{i=1}^N  \langle
0\vert \cos\,  (\theta/2)  +
 i (\sigma_i^x /2)\sin\, (\theta/2)\vert 0\rangle\nonumber\\ &=& [ \cos\,  (\theta/2)]^N \quad
{\longrightarrow}_{N\to\infty}~~ 0\, .
\end{eqnarray} It is easy to verify that the orthogonality in the limit $N\to\infty$ still holds
between excited states involving finite numbers of ``wrong spins'' and hence the Hilbert space of the
system splits into an infinite number of orthogonal Hilbert subspaces built upon the degenerate
ground states labeled by  $ \vec M$. If
$N$ is large but finite,  the orthogonality condition remains approximatively valid if $\theta
>O(1/\sqrt N) $. This is the expected range of quantum fluctuations around a classical configuration
of $N$ aligned spins. One may thus interpret  the stability of a  particular ground state as due to its
{\em classical} nature,  as corroborated by  the   computation of the expectation values. This fact
will be important for the understanding of the difference between   global SBS  and its  local
counterpart.

A feature related to ground state degeneracy under the rotation group is the onset of a normal
mode whose energy vanishes at zero wavevector
$\vec q$. To see this, let us rewrite the Hamiltonian Eq.(\ref{heisenberg}) in terms of  Fourier
components.  Defining
\begin{equation}
\vec S(\vec q) =\frac{1}{\sqrt N}\sum_{i=1}^N \vec S_i\,  e^{i\vec q. \vec r_i}\quad  \quad v(\vec
q)= \frac{1}{ N}\sum_{i\neq j}v_{ij}e^{-i\vec q.(\vec r_i-\vec r_j)}\, ,
\end{equation} Eq.(\ref{heisenberg}) yields, at $h^z=0$,
\begin{equation}
\label {fourier} H_0= -2 \sum_{\vec q} v(\vec q) \vec S(\vec q).\vec S(-\vec q)\, .
\end{equation} Defining $S^\pm= S^x \pm i S^y$  and taking into account  the relation $S^z(\vec q)
\vert 0\rangle =\frac{\sqrt N}{2}\,  \delta_{\vec q, 0} \vert 0\rangle$ one gets, using the
commutation relations of the rotation generators
\begin{equation}
\label{sw} [H_0, S^-(\vec q)] \vert 0\rangle = 2[v(0)-v(\vec q)] S^-(\vec q) \vert 0\rangle\, .
\end{equation} Eq.(\ref{sw}) reveals  a spin wave with energy $\omega$ 
related to the wavevector $\vec q$ by the dispersion relation
\begin{equation}
\label{disperse}
\omega= 2[v(0)-v(\vec q)]\, .
\end{equation} Its energy vanishes in the limit $\vec q \to 0$. This  is a  consequence of the ground
state  degeneracy. The excitation is indeed created  by the operator 
$S^-(\vec q)$ acting on the state $\vert 0\rangle
$, which in the limit $\vec q \to 0$ is proportional to
 generators rotating the degenerate ground states,  and therefore  cannot carry energy. In
relativistic field theory, an excitation whose energy vanishes as 
$\vec q
\to 0$ characterizes a massless mode and   the spin wave may be viewed here as the ``massless''
mode associated with spontaneous broken rotational invariance. It is the ancestor of the
NG  boson that will be discussed in the next section in the context of field theory. Note that
if the external magnetic field
$h^z$ is non-zero, Eq.(\ref{disperse}) gets an additional term in the RHS linear in
$h^z$, and hence a ``mass'' term.

The  effective potential below the Curie point, depicted in Fig.1,   summarizes the essential features of
SBS.  At a given minimum, say, $\vec M=M^z\vec 1_z$, the curvature of the effective potential measures the
inverse susceptibility which determines the energy for infinite wavelength fluctuations, in other words,
the ``mass''. The inverse susceptibility  is zero in directions transverse to the order parameter and
positive in the longitudinal direction. One thus recovers, even at  non-zero temperature, the massless
transverse mode characteristic of  broken continuous symmetry and we learn that there is also a  (possibly
unstable) ``massive'' longitudinal mode which corresponds to fluctuations of the order parameter and which
is present in any spontaneous broken symmetry,  continuous or even discrete. Such generically massive mode
characterize any ordered structure, be it the  broken symmetry phase in
statistical physics, the vacuum of the global SBS in field theory presented in Section 2.2 or of the
Yang-Mills asymmetric phase discussed in Section~3. The ``SBS mass'' of the longitudinal mode
measures the rigidity of the ordered structure.

Consider now some other second order phase transitions.

Superfluidity in He4 occurs when  below a critical temperature a condensate forms out of zero
momentum states of the bosonic atoms. This phenomenon is related to the Bose-Einstein
condensation of a free boson gas, although  interactions reduce the number of particles in the
ground state condensate to a finite fraction of the $N$ atoms of the system. The condensation can be
described by giving to the  creation (or destruction) operator $a_0^+$ at zero momentum a non
vanishing expectation value. The $U(1)$ symmetry of the quantum phase is then spontaneously
broken by selecting a phase. As in ferromagnetism, this results in a degeneracy of the ground state
and in the existence of a concomitant massless mode which here are superfluid sound waves.

 A $U(1)$ broken symmetry also occurs  in superconductivity through condensation of bosonic
Cooper pairs bound states of zero momentum spin singlets
$b_{\vec k}^+=a_{\vec k\uparrow}^+a_{-\vec k\downarrow}^+  $. These are formed because of an
attractive force in the vicinity of the electron Fermi surface induced by phonon exchange. Cooper
pair condensation leads to a gap at the Fermi surface \cite{bcs}.  For neutral superconductors, this
gap hosts a massless mode and one recovers the general features of SBS.  But the presence of the
long-range coulomb interaction modifies the picture. The massless mode disappears by being
absorbed by  electron density oscillations, namely by the longitudinal massive  plasma mode
\cite{anderson, nambu1}. 
 The  penetration depth $1/m_v$ of a magnetic field is a manifestation of a transverse mass
$m_v$. The field is either localized at the boundary in the Meissner effect  if $m_s < O(m_v)$ (Type I
superconductors), or channeled into flux tubes   if $m_s > O(m_v)$ (Type II superconductors). Here
$m_s$ is the SBS mass which measures the rigidity of the condensate. The appearance of these
masses are precursors of the asymmetric Yang-Mills phase presented in Section~3.  
In  superconductivity
 the transverse and longitudinal  masses  are   different and of different dynamical origin. While the
transverse  mass is due to the condensate, the longitudinal one uses the total electron density and
is  also present in the normal phase. In the relativistic theory of Section 3 there will be, due to the 
Lorentz invariance, only one photon mass.

\subsection{Broken continuous symmetry in field theory}

Spontaneous symmetry breaking was introduced in relativistic quantum field theory by Nambu  in
analogy with the BCS theory of superconductivity.   The problem studied by Nambu
\cite{nambu2} and Nambu and Jona-Lasinio \cite{nambujl} is the spontaneous breaking of chiral
symmetry induced by a fermion  condensate $\langle
\bar\psi\psi\rangle \neq 0$. They consider massless fermions interacting  through the  four Fermi
interaction $g [ (\bar \psi\psi )^{2}-(\bar \psi\gamma _{5}\psi )^2]$ that is invariant under the $U(1)$
chiral group transformation $\psi\to \exp (i\gamma_5 \alpha)\psi $. This is a global symmetry as
$\alpha$ is constant in space-time. Although no fermion mass may arise perturbatively, summing up
an infinity of  diagrams  allow the  self-energy to acquire self-consistently a non-zero contribution
from
$\langle
\bar\psi\psi\rangle $.  This yields  a  fermion mass $m$ and  an eigenvalue equation
$g=f(m/\Lambda)$ where $\Lambda$ is a ultraviolet cut-off. The eigenvalue equation  in turn implies
the existence of a massless pseudoscalar mode coupled to the axiovector current.  This  is a
consequence of   the chiral Ward identity relating the axial vector vertex
$\Gamma _{\mu 5}$ to the self energy $\Sigma = A(p^2)\gamma^\mu p_\mu +M(p^2)$ in a chiral
invariant theory,
\begin{equation}
\label{ward} q^\mu\Gamma _{\mu 5}=\gamma _{5}\Sigma (p+\frac{q}{2})+\Sigma(p-\frac{q}{2})
\gamma _{5}\, .
\end{equation} As $q_{\mu }\rightarrow 0$,
 the form factor $A(p^2)$ drops out of Eq.(\ref{ward}) and
\begin{equation}
\label {pseudo}
\lim_{q_{\mu }\rightarrow 0}q_{\mu }\Gamma _{\mu 5}=2M(p^2)\gamma _{5}\quad \quad \Gamma
_{\mu 5}\rightarrow
\frac{2M(p)\gamma _{5}q_{\mu }}{q^{2}}\, .
\end{equation} The pole at $q^2=0$  in Eq.(\ref{pseudo}) signals  the appearance of the pseudoscalar
boson.

The model is not intended to be realistic but sets the scene for more general considerations. The
pion is identified with the massless mode of spontaneous broken chiral invariance. It  gets its tiny
mass (on the hadron scale) from a small explicit breaking of the symmetry, just as a small external
magnetic field
$h^z$ imparts a small gap in the spin wave spectrum. This interpretation of the pion mass
constituted a breakthrough in our understanding of strong interaction physics.

General features of SBS in relativistic quantum field theory were further analyzed by Goldstone,
Salam and Weinberg
\cite{goldstone, gsw}. Here, symmetry is broken by non vanishing  vacuum expectation values of
scalar fields.   The method is designed to exhibit the appearance of a  massless  mode out of the
degenerate  vacuum and does not really depend on the significance of the scalar fields. The latter
could be elementary or represent  collective  variables of   more fundamental fields, as would be
the case in the original Nambu model. Compositeness  affects details of the model considered, such
as the behavior at high momentum transfer and the stability of the SBS massive scalar,  but  not  the
existence of the massless excitations encoded in the degeneracy of the vacuum.
 
Let us  illustrate the occurrence of this massless Nambu-Goldstone  boson in a simple
model of a complex scalar field with $U(1)$ symmetry \cite{goldstone}. The Lagrangian density,
\begin{equation}
\label{global} {\cal L} =\partial ^\mu\phi^*\partial_\mu\phi -V(\phi^*\phi)
\quad\hbox{with} \quad V(\phi^*\phi)= -\mu^2 \phi^*\phi  + \lambda (\phi^*\phi)^2~~,~~\lambda >
0\, ,
\end{equation}  is invariant under the   $U(1)$ group $\phi\to
\displaystyle {e^{i\alpha} \phi}$.  The global $U(1)$ symmetry is broken  by a vacuum expectation
value of the
$\phi$-field given, at the classical level, by the minimum of
$V(\phi^*\phi)$. Writing
$\phi = (\phi_1 + i
\phi_2)/ \sqrt2$, one may choose $\langle
\phi_2\rangle=0$. Hence $ \langle
\phi_1\rangle^2=\mu^2/\lambda $ and we select, say,  the vacuum with 
$\langle
\phi_1\rangle
 $  positive.   The potential
$V(\phi^*\phi)$ is depicted in Fig.2. It is similar to the effective potential below the ferromagnetic
Curie point shown in Fig.1 and leads to similar consequences.

\vskip .5cm
\hskip 2cm\epsfbox{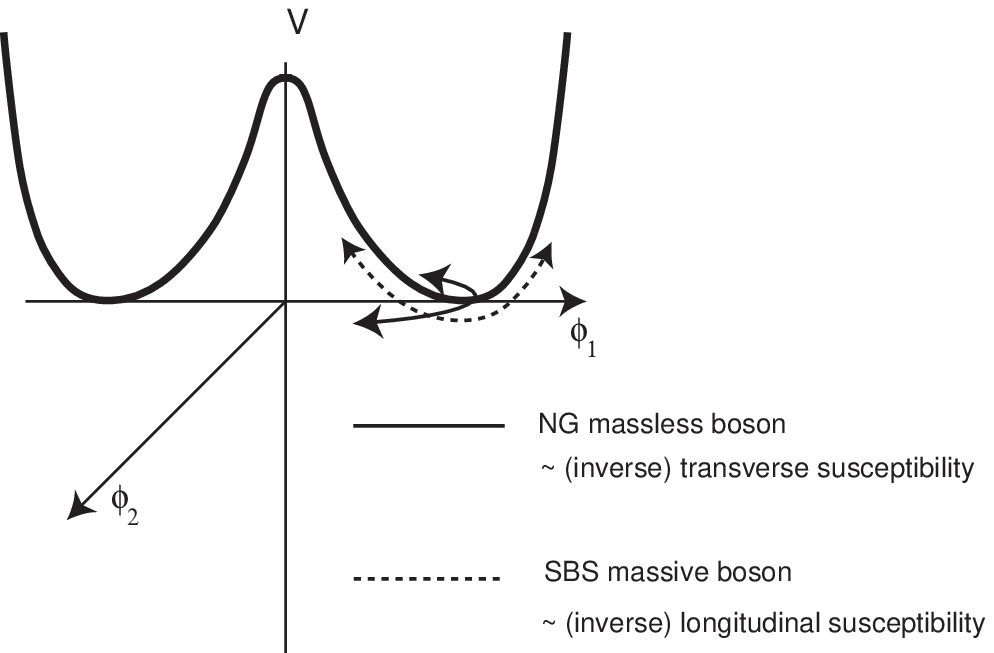}
\begin{quote}
\begin{center}
\baselineskip 10pt {\footnotesize Fig.2.  Spontaneous breaking of a continuous symmetry by 
 scalar fields.}
\end{center}
\end{quote}
\vskip .3cm

In the unbroken vacuum the field $\phi_1$ has negative mass and acquires a positive mass
$2\mu^2$ in the broken vacuum where the field
$\phi_2$ is massless. The latter is the NG boson of broken U(1) symmetry and is the analog of the
massless spin wave mode in ferromagnetism, corresponding to the vanishing of the   inverse 
transverse susceptibility. The massive scalar describes the fluctuations of the order parameter
$\langle\phi_1\rangle$ and is the analog of the  SBS massive  mode in the ordered phase of a many-body
system, encoded in a non-vanishing  inverse longitudinal susceptibility. 

The origin of the massless NG boson is, as in  the ferromagnetism phase, a consequence of the
vacuum degeneracy. The vacuum characterized by the order parameter
$\langle\phi_1\rangle$ is rotated into an equivalent vacuum by an operator proportional to  the
field 
$\phi_2$ at zero space momentum. Such rotation costs no energy and thus the field $\phi_2$ at space
momenta
$\stackrel{\to}{q} =0$ has $q_0=0$  in the equations of motion, and   hence  zero mass.

This Goldstone theorem can be formalized  and generalized by noting that the conserved Noether
current
 $J_\mu= \phi_1\partial_\mu \phi_2 -\phi_2\partial_\mu \phi_1$ gives a charge $ Q =\int  J_0 d^3x$.
The operator $\exp\, (i\alpha Q)$ rotates the vacuum by an angle $\alpha$. In the classical limit, this
charge is, around the chosen vacuum, $ Q =\int  \langle\phi_1\rangle \partial_0 \phi_2 d^3x$. More
generally, 
$\langle[Q, \phi_2]\rangle = i
\langle\phi_1\rangle $ is non zero in the chosen vacuum. This implies that the propagator
$\partial^\mu  \langle T J_\mu(x) ~\phi_2 (x^\prime)\rangle$ cannot vanish at zero four-momentum
$q$ because its integral over space-time is precisely $\langle[Q, \phi_2]\rangle $. Expressing the
propagator in terms of Feynman diagrams we indeed see that  the
$\phi_2$-propagator must have a pole at $q^2=0$. The field $\phi_2 
$ is the massless NG boson.
The proof is immediately extended to  spontaneous  global symmetry breaking of  a
semi-simple   Lie group
$\cal G$ to $\cal H$ . Let $\phi^ A$ be scalar fields spanning a representation of 
$\cal G$ generated by the (antihermitian) matrices
$T^{aAB}$. If the $\cal G$-invariant  potential has  minima
for non vanishing $\phi^ A$,s , the global symmetry is broken  and the vacuum is degenerate under
$\cal G$-rotations. The conserved charges are $ Q^a=\int 
\partial_\mu\phi^ B~ T^{aBA}~\phi^ A ~d^3x$. The propagators of the
fields $\phi^B$ such that $\langle[Q^a,
\phi^B]\rangle = T^{aBA}~\langle\phi^ A\rangle \neq 0$   have a NG pole at
$q^2=0$ and the NG bosons live in the coset $\cal G/H$.

\section{The asymmetric Yang-Mills phase}
\subsection{Global to local symmetry}

The global $U(1)$ symmetry in Eq.(\ref{global})  is extended  to a local one $\phi(x)\to
\displaystyle {e^{i\alpha(x)} \phi(x)}$ by introducing a vector field
$A_\mu(x)$ transforming as  $A_\mu(x)\to A_\mu(x) + (1/ e)
\partial_\mu \alpha(x)$. The  corresponding Lagrangian density is
\begin{equation}
\label{local} {\cal L} =D^\mu\phi^*D_\mu\phi -V(\phi^*\phi) -{1\over4} F_{\mu\nu} F^{\mu\nu} \, ,
\end{equation} with covariant derivative $D_\mu\phi =
\partial_\mu\phi -ieA_\mu
\phi $ and $F_{\mu\nu} = \partial_\mu A_\nu -\partial_\nu A_\mu $.

Local invariance under a semi-simple Lie group $\cal G$ is realized by extending  the Lagrangian 
Eq.(\ref{local}) to incorporate  non-abelian
 Yang-Mills vector fields $A_\mu^a$ 
\begin{eqnarray}
\label{localym} &&{\cal L}_{\cal G} =  ( D ^\mu\phi)^{*A} (D_\mu\phi)^A -V  -\displaystyle { {1\over
4}} F_{\mu\nu}^aF^{a\,\mu\nu}\, ,\\
\label{covariant} &&(D_\mu \phi)^A =\partial_\mu \phi^A -  eA_\mu^a T^{a\, AB }\phi^B\quad \quad
F_{\mu\nu}^a =\partial_\mu A_\nu^a -\partial_\nu A_\mu^a -e f^{abc}A_\mu^b A_\nu^c\,
.\qquad
\end{eqnarray} Here, $\phi^A$ belongs to the representation of
$\cal G$ generated by
$T^{a\,AB }$ and the potential $V$ is invariant under  $\cal G$.

The local abelian or non-abelian gauge invariance of Yang-Mills theory hinges {\em apparently}  upon
the massless character of the gauge  fields $A_\mu$, hence on the long-range character of the forces
they transmit,  as the addition of a mass term for $A_\mu$ in the Lagrangian Eq.(\ref{local}) or
(\ref{localym})  destroys gauge invariance. But  short-range forces such as the weak interaction
forces, 
  seem to be as fundamental as the electromagnetic ones despite the apparent departure from
exact  conservation laws. To reach a basic description of such forces one is  tempted to link this
fact  to gauge fields masses
 arising from spontaneous broken  symmetry.  However the problem of  SBS is  different for global
and for local symmetries.

To pinpoint the difference, let us break the symmetries explicitly. To the Lagrangian Eq.(\ref
{global}) we add, in analogy with the  magnetic field $h^z$ in Eq.(\ref{heisenberg}), the term
\begin{equation}
\label{breakglobal}
\phi h^* +\phi^* h\, ,
\end{equation} where $h , h^*$ are constant in space time. Let us take $h$ real. The presence of the 
field $h$ breaks explicitly the global $U(1)$ symmetry and the field
$\phi_1$  develops an expectation value. When $h\to 0$, the symmetry of the action is restored but,
when the symmetry is broken by a minimum of $V(\phi\phi^*)$ at $\vert\phi\vert   \neq 0$, we still
have
$\langle
\phi_1\rangle \neq 0$. The tiny
$h$-field  has simply picked up one of the degenerate vacua  in perfect analogy with the
infinitesimal magnetic field which orients the magnetization of a ferromagnet. As in the latter, the
chosen vacuum is stable  because it is defined by a classical configuration of the fields  and the
Hilbert space breaks up into an infinite set of disjoint spaces.
 The degeneracy of the vacuum can be put into evidence by changing the phase of
$h$; in this way, we can  reach in the limit
$h\to 0$ any  $U(1)$ rotated vacuum. 

When the symmetry is extended from global to local, one can still break the gauge symmetry by an
external ``magnetic'' field, say a mass term. However in the  zero mass limit,  no preferred vacuum
exists.  In contradistinction to the global symmetry case,  no energy is needed to change the {\em
relative} orientation of neighboring ``spins'', that is of neighboring configurations of  the scalar fields
in group space.  As a consequence,  no  classical configuration is available to  protect a degenerate
vacuum against quantum fluctuations.

This fact has two related consequences.

First, the vacuum is generically non degenerate and points in no particular direction in group 
space\footnote{Note that for global symmetry breaking, one can always choose a linear combination
of degenerate vacua which is invariant under, say, the
$U(1)$ symmetry. This  choice  has no observable consequences because of the splitting into
orthogonal Hilbert spaces.}.
In this sense, local gauge symmetry cannot be spontaneously broken\footnote{For a detailed
 proof, see reference
\cite{elitzur}.}. 

As a consequence, there cannot be massless NG bosons. These correspond to relative orientation of
neighboring ``spins'' and are now simply gauge transformations. A formal proof of the failure of the
Goldstone theorem in presence of gauge fields,  in relativistic quantum field theory, was given by
Higgs\cite{higgs1}.

 Recalling that the explicit presence of a gauge vector mass
 breaks gauge invariance, we are thus faced with a dilemma.  How can  gauge fields acquire mass
 without breaking the local symmetry?

In perturbation theory, gauge invariant quantities are evaluated by choosing a particular gauge. One 
imposes the gauge condition by adding to the action a gauge fixing term and the corresponding Fadeev-Popov
ghosts, and  gauge invariance is ensured by summing over subsets of graphs satisfying the Ward Identities.

Consider the Yang-Mills theory defined by the Lagrangian Eq.(\ref{localym}). To exhibit the
similarities and the differences between spontaneous breaking of a global symmetry and its local
symmetry counterpart, it is convenient to choose a gauge
which preserves Lorentz invariance and a residual global $\cal G$ symmetry. This can be achieved by
adding to the Lagrangian a gauge fixing term $(2\eta)^{-1} \partial_\mu A^{a\mu} \, \partial_\nu
A^{a\, \nu} $. The gauge parameter $\eta$ is  arbitrary and is not observable.

In such gauges, the global symmetry can  be spontaneously broken for suitable potential $V$,
by  non zero expectation values $\langle
\phi^A\rangle$ of scalar fields. In Fig.3 we have represented   motions of this parameter in the
spatial $q$-direction and  in  a direction $B$ of the coset space $\cal G/H$ where $\cal H$ is the
unbroken subgroup.  Fig.3a
pictures the spontaneously broken vacuum of the  gauge fixed Lagrangian. Fig.3b and Fig.3c mimic
 motions in the coset with decreasing wavelength $\lambda$.  Clearly, as
$\lambda\to \infty$, such motions can only induce global rotations in the internal space.
In absence of gauge   fields,  they would give rise, as in  spontaneously broken  global continuous
symmetries, to  massless NG modes generating the coset in the limit $\lambda=\infty$. In a gauge
theory,  transverse fluctuations of
$\langle\phi^A\rangle$ are just   local rotations in the internal space and
 are  unobservable gauge motions.  Hence  the would-be NG bosons induce only
gauge transformations and their  excitations disappear from the physical  spectrum. 

\vskip .5cm
\hskip 1.5cm
\epsfbox{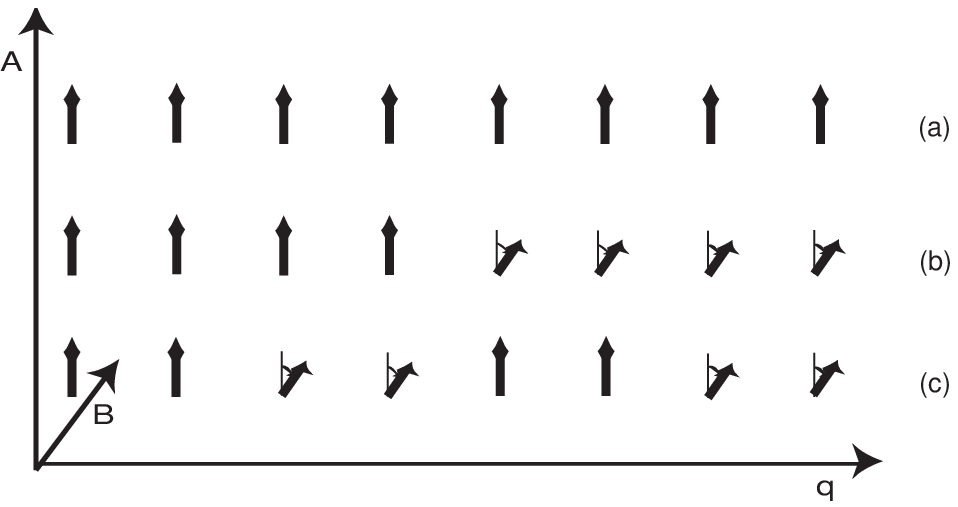}
\begin{quote}
\begin{center}
\baselineskip 10pt {\footnotesize Fig.3.  The disappearance of the massless NG boson in a gauge
theory.}
\end{center}
\end{quote}
\vskip .2cm

But what makes local internal space rotations    unobservable in a
gauge theory is precisely the fact that they can be absorbed  by the
Yang-Mills fields. The absorption of the  NG fields renders massive the  gauge fields
living in the coset $\cal G/H$  by transferring to them their degrees of
freedom which become  longitudinal polarizations. 

We shall see in the next sections how these  considerations  are  realized in relativistic quantum
field  theory and give rise to   vector masses  in the coset
$\cal G/ H$, leaving long-range forces only in a subgroup $\cal H$ of $\cal G$.  Despite the unbroken
local symmetry,  the group $\cal G$ appears   broken to its
subgroup
$\cal H$ in the asymptotic state description of field theory,  and I shall therefore often term  SBS or
asymmetric  such a Yang-Mills phase. The onset of SBS will be described in detail mostly in lowest
order perturbation theory around the self-consistent vacuum, both in the field-theoretic
\cite{eb} and in the equation of motion \cite{higgs2} formulations.  This contains already the basic
ingredients of the phenomenon and comparison between the two methods gives some insight on
the renormalization issue.

\subsection{The field theoretic approach}

{\it $\alpha$) Breaking by  scalar fields}
\medskip

Let us first examine the abelian case as realized by the complex scalar field $\phi$  exemplified in
Eq.(\ref {local}).

In the covariant gauges, the free propagator of the  field
$A_\mu$ is
\begin{equation}
\label {dabelian} D_{\mu\nu}^0 ={g_{\mu\nu}-q_\mu q_\nu /q^2\over q^2} + \eta {q_\mu q_\nu/q^2
\over q^2}\, ,
\end{equation} where $\eta$ is the  gauge parameter. 

In absence of symmetry breaking, the lowest order contribution to the self-energy, arising from the
covariant derivative terms in Eq.(\ref{local}), is given by the one-loop diagrams of Fig.4. The
self-energy (suitably regularized) takes the form of a polarization tensor 
\begin{equation}
\label{pabelian}
\Pi_{\mu\nu}= (g_{\mu\nu} q^2-q_\mu q_\nu)~ \Pi(q^2)\, ,
\end{equation} where the scalar polarization $\Pi(q^2)$ is regular at
$q^2=0$, leading to the gauge  field propagator
\begin{equation} D_{\mu\nu} ={g_{\mu\nu}-q_\mu q_\nu /q^2\over q^2[1-
\Pi(q^2)]} +\eta {q_\mu q_\nu/q^2\over q^2}\, . 
\end{equation} The  polarization tensor in Eq.(\ref{pabelian}) is transverse and hence does not affect
the gauge parameter 
$\eta$.  The transversality of the polarization tensor reflects the gauge invariance of the
theory\footnote{The transversality of polarization tensors is a consequence of the Ward Identities
alluded to in the preceding section.} and, as we shall see below, the regularity of the polarization
scalar signals the absence of symmetry breaking. This guarantees that the  
$A_\mu$-field remains massless.

\vskip 1cm
\hskip 1.5cm
\epsfbox{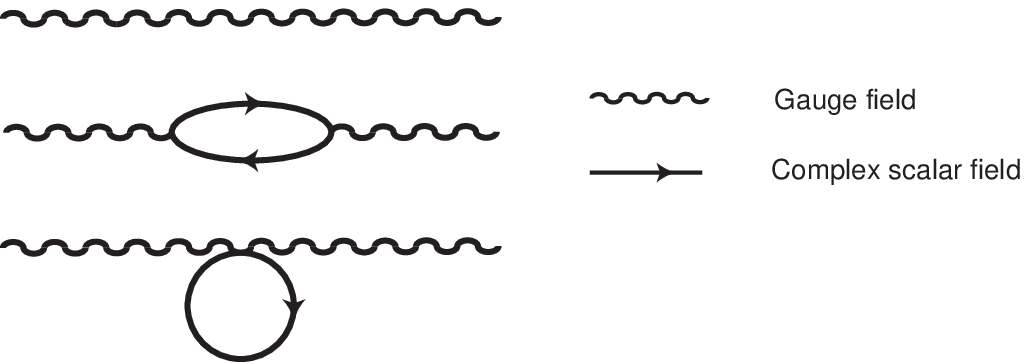} 

\begin{quote}
\begin{center}
\baselineskip 10pt {\footnotesize Fig.4.  Lowest order vacuum polarization graphs in absence of SBS.

Abelian gauge theory.}

\end{center}
\end{quote}
\vskip .2cm

Symmetry breaking adds tadpole diagrams to the previous ones. To see this write
\begin{equation}
\label{vacuum}
\phi ={1\over \sqrt 2}(\phi_1 + i\phi_2)~~~
\langle
\phi_1\rangle
\neq 0\, .
\end{equation} The scalar  field $\phi_1$, whose expectation value plays  the role of  an order
parameter in the gauge considered and whose fluctuations have a gauge invariant SBS mass, is often
called the Higgs field and its  fluctuations the Higgs boson.  This massive mode is  not a specific
property of the BEH mechanism but  is a necessary concomitant of {\em any} SBS  structured vacuum,
as pointed out in Section 2.1.  The would-be NG-field is
$\phi_2$. The additional diagrams are depicted in Fig.5.
\vskip .5cm
\hskip 2cm
\epsfbox{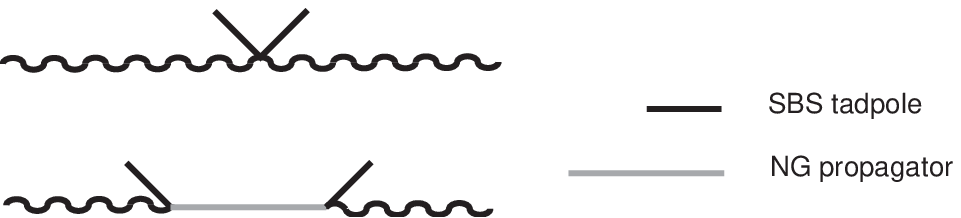}
\begin{quote}
\begin{center}
 {\footnotesize Fig.5.  Tadpole graphs  of SBS. Abelian gauge theory.}
\end{center}
\end{quote}
\vskip .3cm
\noindent The polarization scalar $\Pi(q^2)$ in Eq.(\ref{pabelian})   acquires a pole from the tadpole
contribution
\begin{equation}
\label{pamass}
\Pi(q^2)={e^2\langle \phi_1\rangle^2\over q^2}\, ,
\end{equation} and, in lowest order perturbation theory,  the gauge  field propagator becomes
\begin{equation}
\label{damass} D_{\mu\nu} ={g_{\mu\nu}-q_\mu q_\nu/q^2
\over q^2- \mu^2}  +\eta {q_\mu q_\nu/q^2\over q^2}\, , 
\end{equation} which shows that the $A_\mu$-field gets a mass
\begin{equation}
\label{massa}
\mu^2=e^2 \langle
\phi_1\rangle^2\, .
\end{equation}

The generalization of Eqs.(\ref{pabelian}) and (\ref{pamass}) to the non abelian case described by
the action Eq.(\ref{localym}) is straightforward. One gets from the graphs depicted in Fig.6,

\vskip .5cm
\hskip 3.5cm\epsfbox{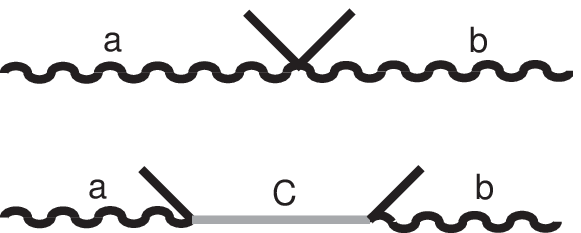}
\begin{quote}
\begin{center}
\baselineskip 10pt {\footnotesize Fig.6.  Tadpole graphs  of SBS. Non-abelian gauge theory.}
\end{center}
\end{quote}
\begin{eqnarray}
\label{pnamass} &&\Pi_{\mu\nu}^{ab}= (g_{\mu\nu}q^2-q_\mu q_\nu )
\Pi^{ab}(q^2)\ , \hskip 3cm\\
\label{pol} && 
\Pi^{ab}(q^2)={e^2\langle \phi^{*B}\rangle T^{*a\,BC}T^{b\,CA}\langle\phi^A\rangle \over q^2}\, .
\end{eqnarray} 
Eq.(\ref{pol})  defines the mass matrix
\begin{equation}
\label{massna}
(\mu^2)^{ab} = e^2\langle \phi^{*B}\rangle T^{*a\,BC}T^{b\,CA}\langle\phi^A\rangle\, . 
\end{equation} In terms of its non-zero eigenvalues
$(\mu^2)^a$,  the propagators of the massive gauge vectors take the same form as
Eq.(\ref{damass}),
\begin{equation}
\label{dnamass} D_{\mu\nu}^a  ={g_{\mu\nu}-q_\mu q_\nu/q^2
\over q^2- (\mu^2)^a}  +\eta {q_\mu q_\nu/q^2\over q^2}\, . 
\end{equation} 

The gauge invariance is expressed, as it was in absence of symmetry breaking, through the
transversality of the polarization tensors Eqs.(\ref{pabelian}) and (\ref{pnamass}).  The singular 
$1/q^2$ contributions to
 the polarization scalars  Eqs.(\ref{pamass}) and (\ref{pol}) preserve transversality and
 yield  gauge invariant masses for the gauge bosons. They stem  from the long-range  NG boson
fields.  The latter are, as such,  unobservable gauge terms but
their absorption in the gauge field propagators transfers  the degrees of freedom of the would-be
NG bosons to the third degree of polarization  of the massive vectors. Indeed, on the mass shell
$q^2= (\mu^2)^a$, one easily verifies that the numerator in the transverse propagator in Eq.(\ref
{dnamass}) is
\begin{equation} g_{\mu\nu}-{q_\mu q_\nu\over q^2}=\sum_{\lambda=1}^3
e_\mu^{(\lambda)}.e_\nu^{(\lambda)}\qquad q^2=
(\mu^2)^a 
\, ,
\end{equation} where the $e_\mu^{(\lambda)}$ are  three polarization vectors 
orthonormal in the rest frame of the particle.  

In this way, the would-be NG bosons generate massive propagators for the gauge fields in $\cal
G/H$. Long-range forces only survive in the subgroup
$\cal H$ of
$\cal G$ which leaves invariant the non vanishing expectation values
$\langle\phi^A\rangle$.

Note that  the explicit form of the scalar potential $V$  does not enter the
computation of  gauge field propagators which depend only on the expectation values  at its
minimum. This is because  trilinear terms arising from  covariant derivatives, which yields the
second graphs of Fig.5 and Fig.6,  can only couple the tadpoles to  other scalar fields  through group
rotations and hence couple them only  to the would-be NG bosons. These are the eigenvectors with
zero eigenvalue of the scalar mass matrix given by the quadratic term in the expansion of the
potential
$V$ around its minimum. Hence the massive scalars decouple from the tadpoles at the tree level
considered above.  

\medskip
 {\it $\beta$) Dynamical symmetry breaking}
\medskip

The symmetry breaking giving mass to gauge vector bosons may arise from the fermion condensate
breaking chiral symmetry. This is illustrated by the following chiral invariant Lagrangian
\begin{equation} {\cal L} = {\cal L}_0^F -e_V\,\bar\psi\gamma_\mu\psi V_\mu 
-e_A\,\bar\psi\gamma_\mu\gamma_5\psi A_\mu  -  {1\over 4} F_{\mu\nu}F^{\mu\nu}(V) - {1\over 4}
F_{\mu\nu}F^{\mu\nu}(A)\, .
\end{equation} Here $F_{\mu\nu}(V)$ and $F_{\mu\nu}(A)$ are abelian field strength for
$U(1)\times U(1)$ symmetry.  Chiral anomalies are eventually canceled by adding in the required
additional fermions.

As in global SBS, the Ward identity for the chiral current Eq.(\ref {ward})
 shows that if  the fermion self-energy
$\gamma^\mu p_\mu A(p^2) - M(p^2)
$ acquires  a non vanishing
$M(p^2)
$ term, thus a dynamical mass, the axial vertex
$\Gamma_{\mu 5}$ develops a pole at $q^2=0$.  In leading order in
$q$, we get as in Eq.(\ref{pseudo})
\begin{equation}
\label{vertex}
\Gamma_{\mu 5}\, {\rightarrow}\, 2M(p)\,\gamma_5 {q_\mu\over q^2}
\, .
\end{equation}

\vskip .5cm
\hskip 2cm
\epsfbox{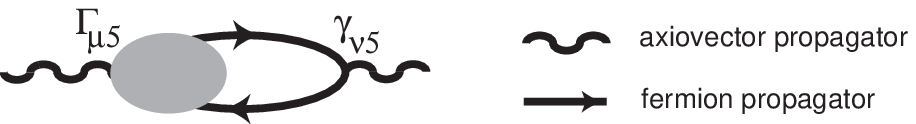}
\begin{quote}
\begin{center}
\baselineskip 10pt {\footnotesize Fig.7.  Dynamical SBS from fermion condensate.}
\end{center}
\end{quote}
\vskip .3cm

The pole in the vertex function induces a pole in the suitably regularized gauge invariant
polarization tensor
$\Pi_{\mu\nu}^{(A)}$ of the axial vector field $A_\mu$ depicted in Fig.7 
\begin{equation}
\label{poldyn}
 \Pi_{\mu\nu}^{(A)}=  e_A^2 (g_{\mu\nu}q^2-q_\mu q_\nu )\Pi^{(A)}(q^2)\, ,
\end{equation}  with
\begin{equation}
\label{pdyn}
\lim_{q^2\to 0} q^2\Pi^{(A)}(q^2) =\mu^2
\neq 0\, .
\end{equation} The field $A_\mu$  acquires in this approximation\footnote {The validity of the
approximation, and in fact of the dynamical approach, rests on the high momentum  behavior of the
fermion self energy, but this problem will not be discussed here.} a gauge invariant mass
$\mu$.

This  example illustrates the fact that the transversality of the polarization tensor used in the
quantum field theoretic approach to mass generation  is a consequence of a Ward identity.  This is
true whether vector masses  arise through fundamental  scalar   or through fermion condensate.  
The generation of gauge invariant masses is therefore not contingent upon the ``tree
approximation'' used to get the propagators Eqs.(\ref{damass}) and (\ref{dnamass}). It is a
consequence of     the
$1/q^2$ singularity in the vacuum polarization scalars Eqs.(\ref{pamass}), (\ref{pnamass}) or
(\ref{pdyn} ) which comes from the would-be NG boson contribution.

\subsection{The equation of motion formulation}
 
The BEH mechanism  can be understood in terms of  equations of motions which illustrate nicely the
fate of the NG bosons. This is shown below for the abelian case described by the action
Eq.(\ref{local}).

Taking as in Eq.(\ref{vacuum}), the expectation value of the  scalar  field to be
$\langle\phi_1\rangle$, and expanding the NG field $\phi_2$ to first order, one gets from the action
Eq.(\ref{local}) the classical equations of motion 
\begin{eqnarray}
\label{first} &&\partial^\mu \{\partial_\mu  \phi_2 -e\langle
\phi_1\rangle A_\mu\} =0\, ,\\
\label{second} &&\partial_\nu F^{\mu\nu}= e \langle
\phi_1\rangle\{\partial^\mu \phi_2 - e\langle
\phi_1\rangle A^\mu\}\, .
\end{eqnarray} 
 Defining
\begin{equation}
\label{hvector}
 B_\mu = A_\mu - {1\over e\langle
\phi_1\rangle}\partial_\mu 
\phi_2~~~\hbox{and}~~~G_{\mu\nu} =
\partial_\mu B_\nu -\partial_\nu B_\mu = F_{\mu\nu}\, ,
\end{equation} one gets
\begin{equation}
\label{vector}
\partial_\mu B^\mu =0 \quad \quad \partial_\nu G^{\mu\nu} + e^2 
\langle
\phi_1\rangle^2 B^\mu=0\, . 
\end{equation} Eq.(\ref{vector}) shows that $B_\mu$ is a massive vector field with mass squared $
e^2 \langle\phi_1\rangle^2 $ in accordance with Eq.(\ref{massa}). As pointed out in the previous
section, the vector boson mass
 does not depend  explicitly on the scalar potential, but only on the value of
$\langle\phi_1\rangle$ at its minimum.

The value of $\langle\phi_1\rangle$ and the mass of the massive  scalar boson are determined by
the potential and are of course not affected by the gauging.  For the  potential Eq.(\ref{global}) one
recovers from the equation of motion for the
 massive   scalar
\begin{equation}
\left\{\partial^2 -  V^{\prime\prime}(\langle\phi_1\rangle)\right\}\delta\phi_1=0\, , 
\end{equation}
 the  mass
$2\mu^2$  using
$\langle\phi_1\rangle^2 =\mu^2/\lambda$. 

In this formulation,    we see clearly from Eq.(\ref {hvector}) how the NG boson is absorbed into a
redefined massive vector field.  The disappearance of the NG boson was  further analyzed in
reference \cite {higgs3}. In the gauge defined by Eq.(\ref{vector}),  the  field
$B_\mu$, which contains only the physical degrees of freedom of the massive vector, does appear.
This  is the ``unitary gauge'' of the theory. In contradistinction,  the field theoretic approach
introduces a spurious
$1/q^2$ pole in the polarization Eq.(\ref{pamass}), which is not observable. The comparison between
these two different approaches to massive gauge vector boson masses
contains the germ of the renormalizability of the BEH mechanism, as will now be discussed.

\subsection{The renormalization issue}

The massive vector propagator Eq.(\ref{dnamass}) differs from a conventional free massive
propagator in two respects. First the presence of the unobservable longitudinal term reflects the
arbitrariness of the gauge  parameter
$\eta$. Second the NG pole at $q^2 =0$ in the transverse projector
$g_{\mu\nu}-q_\mu q_\nu / q^2$ is unconventional. Its significance is made clear by expressing the
propagator of the $A_\mu$ field in Eq.(\ref{dnamass}) as (putting
$\eta$ to zero)
\begin{equation}
\label{dnamass2} D_{\mu\nu}^a
\equiv {g_{\mu\nu}-q_\mu q_\nu/q^2
\over q^2- (\mu^2)^a}  ={g_{\mu\nu}-q_\mu q_\nu/(\mu^2)^a
\over q^2- (\mu^2)^a}  +{1\over (\mu^2)^a} {q_\mu q_\nu
\over q^2}\, .
\end{equation} The first term in the right hand side of Eq.(\ref{dnamass2}) is the conventional
massive vector propagator. It may be viewed as the (non-abelian generalization of the) free
propagator of the
$B_\mu$-field defined in Eq.(\ref{hvector}) while  the second term is  a pure gauge propagator due
to the NG boson ($[1/ e\langle
\phi_1\rangle]\partial_\mu 
\phi_2$  in Eq.(\ref{hvector})~) which converts the gauge field $A_\mu$  into the massive vector field
$B_\mu$.

The propagator Eq.(\ref{dnamass}) which appeared in the field theoretic approach contains thus, in
the covariant gauges, the transverse projector
$g_{\mu\nu}- q_\mu q_\nu/q^2$ in the numerator of the massive gauge
 field $A_\mu^a$ propagator.   This is in sharp contradistinction to the numerator
$g_{\mu\nu}-q_\mu q_\nu/(\mu^2)^a$ characteristic of the conventional massive vector field $B_\mu$
propagator. It is the transversality of  the polarization tensor in covariant gauges, which led in the
tree approximation to the transverse projector in Eq.(\ref{dnamass}). As  mentioned above, the
transversality of the polarization tensor is a consequence of a Ward identity and  therefore does not
rely on the tree approximation. This fact is already clear from the dynamical example
Eq.(\ref{poldyn}) but was proven in more general terms in a subsequent publication\footnote{The
proof given in reference
\cite{be2} was not complete because closed Yang-Mills loops, which would have required the
introduction of Fadeev-Popov ghosts were not included.} \cite{be2}. The importance of this fact is
that  transversality  in covariant gauges determines the power counting of
irreducible diagrams. It is then straightforward to verify that the  quantum  field theory formulation
has the required power counting for a renormalizable field theory. On
 this basis it was  suggested that it indeed was renormalizable
\cite{be2}. 

However power counting is not enough to prove the renormalizability of a theory with local gauge
invariance. In addition, to be consistent, the theory must  also be  unitary, a fact which is not
apparent in  ``renormalizable'' covariant gauges  but is  manifest in the ``unitary gauge'' defined in
the free theory by the
$B_\mu$-field introduced in Eq.(\ref{hvector}).  In the unitary gauge  however, 
power counting requirements fail. The equivalence  between the
 $A_\mu$ and  $B_\mu$ free propagators, {\em which is only true in a gauge invariant  theory} where
their difference is the unobservable NG propagator appearing in Eq.(\ref{dnamass2}), is a clue of the
consistency  of the BEH theory.  It is of course a much harder and subtler affair to proof that the
full interacting theory is both renormalizable and unitary. This was  achieved in the work of 't Hooft
and Veltman~\cite{renorm}, which thereby established the  consistency  of the BEH mechanism. 

\section{The unification paradigm}

I first review very briefly the basic elements of the electroweak theory,  one of the most brilliant
achievements of the twentieth century. Its remarkable success played an important role in the
further quest for unification which has become  a paradigm in most of contemporary
research on fundamental interactions. 

In the electroweak theory, the gauge group is taken to be
$SU(2)\times U(1)$ with corresponding generators and coupling constants
$gA_\mu^aT^a$ and $g^\prime B_\mu Y^\prime $. The $SU(2)$ acts on left-handed fermions only. 
The scalar field $\phi$ is a doublet of $SU(2)$ and its
$U(1)$ charge is $Y^\prime = 1/2$.  Breaking is  characterized by $
\langle\phi\rangle={1/\sqrt2}\  \{ 0, v\} $ and $Q=T^3 + Y^\prime$ generates the  unbroken
subgroup.  $Q$ is identified with the electromagnetic charge
operator. The only residual massless gauge boson is the photon
and the electric charge $e$ is usually expressed in terms of the mixing angle
$\theta$ as $g=e/\sin\theta  , g^\prime=e/\cos\theta $.

Using Eqs.(\ref{massa}) and (\ref{massna}) one gets the mass matrix
\begin{center}
$\vert\mu^2\vert$=$\displaystyle{{v^2\over 4}} ~\begin{array}{|cccc|} g^2&0&0&0\\ 0&g^2&0&0\\
0&0&g^{\prime 2}& -g g^\prime\\ 0&0&-g g^\prime & g^2
\end{array}$
\end{center} whose diagonalization yields the eigenvalues
\begin{equation} M^2_{W^+}={v^2\over 4}g^2\qquad M^2_{W^-}={v^2\over
4}g^2\qquad M^2_Z={v^2\over 4}~(g^{\prime 2}+ g^2)\qquad M^2_A= 0 \, . 
\end{equation} This permits to relate $v$ to  the the four Fermi coupling $G$, namely 
$v^2= ({\sqrt 2} G)^{-1}$. 

Although the electroweak theory has been amply verified by experiment, the existence of the
massive scalar boson has, as yet, not been confirmed. It should be noted that its physics  is, as
previously discussed, more sensitive to the dynamical assumptions of the model than the massive
vectors
$W^\pm$ and 
$Z$, be it a genuine elementary field or a manifestation of a composite due to a more elaborate
mechanism. Observation of its mass and width is of particular interest for further understanding of
the mechanism at work. 

The discovery that confinement could be found in the strong coupling limit of quantum
chromodynamics based on the ``color'' gauge group
$SU(3)$ led to tentative Grand Unification schemes where electroweak  and strong interaction could
be unified in a simple gauge group ${\cal G}$ containing $
 SU(2)\times U(1)\times SU(3)$  \cite{gqw}. Breaking occurs through vacuum expectation values of
scalar fields and unification is apparent at high energies because, while the renormalization
group makes the small gauge coupling of
$U(1)$  increase logarithmically with the energy scale, the converse is true for the asymptotically
free non abelian gauge groups. 

Originally the BEH mechanism was conceived to unify the theoretical description of long-range and
short-range forces. The success of the electroweak theory made the mechanism a candidate for
further unification. Grand unification schemes,  where the scale of unification is pushed close to the
scale of quantum gravity effects, strengthen the believe in a still larger unification that would
include gravity. This trend towards  unification received a further impulse from the developments of
string theory  and from its connection  with eleven-dimensional supergravity. The latter is then
often viewed as a classical limit of a hypothetical M-theory into which all perturbative string
theories would merge to yield a comprehensive theory of ``all'' interactions.

Such vision may be premature.  Quite apart from obvious
philosophical questions raised by a ``theory of everything'' formulated in the present framework of
theoretical physics, the transition from perturbative string theory to its M-theory generalization
hitherto  stumbles on the treatment of non perturbative gravity. This might well  be a hint
that new conceptual elements have to be found to cope with the relation between  gravity and
quantum theory and which might not be directly related to the unification program.

\section{Further developments : conceptual issues}

Aside from, or part of,  the unification program,  the BEH mechanism has put into evidence concepts
which may have a profound impact on further research. One of the richest sources of such concepts
is the discovery by 't Hooft and Polyakov of regular monopoles in non abelian gauge theories. I shall
review the underlying features which are present in {\em all } semi-simple Lie groups and stress
their implications. Also of  interest is the geometrical interpretation of the mechanism in the
context of the string theory approach.

\subsection{\bf  Monopoles,  electromagnetic duality, confinement} 

In electromagnetism, monopoles can be included at the expense of introducing a Dirac string \cite
{dirac}. The latter creates a singular potential along a string  ending at the monopole.   

\hskip 4cm \epsfbox{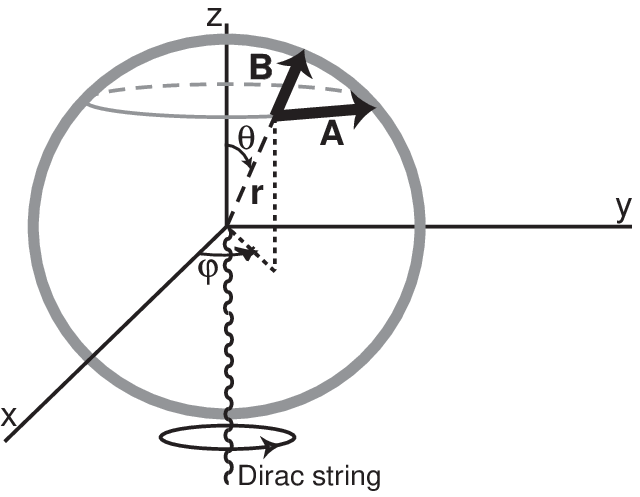}
\begin{quote}
\begin{center}
\baselineskip 10pt {\footnotesize Fig.8. Dirac monopole in abelian gauge theory.}
\end{center}
\end{quote}
\vskip .2cm
\noindent For instance to describe a point-like monopole located at
$\vec r =0 $, one can take the line-singular potential
\begin{equation}
\label{abmono}
\vec A= {g\over 4\pi} ( 1-\cos \theta) 
\vec \nabla \varphi \qquad \vec B=\vec\nabla\wedge \vec A\, .
\end{equation}  This potential has a singularity along the negative $z$-axis
$(\theta =\pi) $ where the string has been put (see Fig.8). The unobservability of the string implies
that its fictitious flux be quantized according to the Dirac condition
\begin{equation}
\label{dirac} eg =2\pi n\ \quad n\in Z \, .
\end{equation} 

In contradistinction to the string in the $U(1)$ theory, the Dirac string in non abelian gauge groups
can be removed  by a gauge transformation for well defined magnetic charge quantization  encoded
in the global structure of the group.  An  $SO(3)$ regular monopole was obtained by 't~Hooft and
Polyakov \cite {hp} by breaking the symmetry to $U(1)$ by scalar fields $\phi^a$ belonging to the
adjoint representation. In a point-singular limit corresponding to vanishing matter current, denoting
by
$A^{i\, a}$ the space components of the isovector fields $\vec A^a$,  it is given by
\begin{equation}
\label{regular} A^{i\,a}={g\over 4\pi} \epsilon^{i ja} { r^j\over r^2}\qquad  \phi^a=\frac{r^a}{r}F
\qquad eg =4\pi \, .
\end{equation}
 Here  the structure constants
in the Yang-Mills  Lagrangian are normed to $\epsilon^{abc}$ and the constant $F$ is fixed by the
minimum of the scalar potential. The potential
$\vec A^a$ in Eq.(\ref{regular})  is the ``spherical'' gauge transformed of the solution     given in the
``abelian'' gauge by Eq.(\ref{abmono}) with  
$\vec A=\vec A^3\, (\vec A^2=\vec A^1=0)$ and by a scalar
isovector 
$\phi^3\, (\phi^1=\phi^2=0)$   constant in space.  In performing the gauge transformation to the
spherical gauge the Dirac string has been removed. The point singularity is smeared in solutions
with non vanishing matter current to yield a   topologically stable
't~Hooft-Polyakov regular monopole
\cite {hp}. 

This analysis  can be extended to all semi-simple Lie  groups $\cal G$
\cite {all}. For a general Lie groups $\cal G$, the
possibility of gauging out the Dirac string depends on the global properties of $\cal G$. Namely, the
map of an infinitesimal curve surrounding the Dirac string into $\cal G$ must be a curve continuously
deformable to zero\footnote{Alternatively, one may require that the Wu and Yang potentials
\cite {wy}  be gauge equivalent in the overlapping region
\cite{all}.}. For sake of brevity I
 limit here the discussion of this condition to   Yang-Mills theories for simple Lie group with  scalar
matter fields belonging to  the adjoint representation of the group. 

The full Lagrangian Eqs.(\ref{localym}),  (\ref{covariant}) is invariant under the group   ${\cal
G_A}=\widetilde{\cal G}/Z$ where
$\widetilde{\cal G}$ is the universal covering group of  the adjoint group
$\cal G_A$ and $Z$ its center.  Let the potential be such that the symmetry breaks  to ${\cal
H}={\cal T}$ where ${\cal T}$ is a  maximal abelian subgroup of 
 $\cal G_A$. It is easily seen that the Lagrangian Eq.(\ref{localym}) admit   line-singular  solutions 
with the gauge fields $A^a_\mu (x)$ in abelian configurations of the type Eq.(\ref{abmono}), namely
\begin{equation}
\label{nabmono}
 \vec  A^a= {g^a\over 4\pi} ( 1-\cos \theta) 
\vec \nabla \varphi\qquad \phi^a= {\rm constant}\qquad a\in {\cal T} \, .
\end{equation}  The condition for the string to be unobservable is
\begin{equation}
\label{unobservable}
\exp \left(i eg^a {\bf t}_a^{(\cal G_A)}\right)=1\, ,
\end{equation} where ${\bf t}_a^{(\cal G_A)}$ are  abelian generators in a faithful representation of
$\cal G_A$. The condition Eq.(\ref{unobservable}) expresses that a closed curve in space is mapped
onto a closed curve in the group space of $\cal G_A$ starting and ending at the unit element. If this
curve can be continuously shrunk to zero, the Dirac string can be removed, leaving a point-singular
solution. This implies
\begin{equation}
\label{nostring}
\exp \left(i eg^a {\bf t}_a^{(\tilde {\cal G)}}\right)=1\, ,
\end{equation} where ${\bf t}_a^{(\tilde {\cal G)}}$ is a  faithful representation of $\tilde{\cal G}$. 
Except for $G_2, F_4$ and
$E_8$ that have $Z=1 $, Eq.(\ref{nostring}) yields a more stringent condition than
Eq.(\ref{unobservable}). The closed curves in
$\cal G_A$ which are homotopic to zero are only those which correspond to the trivial element of
$Z$, or equivalently to closed curves in
$\tilde{\cal G}$. 

All  eigenvalues of ${\bf t}_a^{(\tilde {\cal G)}}$ are vectors $\vec m$ of the weight lattice
$\Lambda_W$ of
$\tilde {\cal G}$ of which the root lattice  $\Lambda_R$ generated by the simple roots
$\vec\alpha_i$ is a sublattice. The simple roots are normalized by the structure constants used in
the Yang-Mills Lagrangian, which generate the adjoint representation of the group. The  lattice
$\Lambda_R^\vee$ generated by the coroots $\vec\alpha_i^\vee=
2\vec\alpha_i/\vec\alpha_i.\vec\alpha_i$ of the simple roots $\vec\alpha_i$ is dual to
$\Lambda_W$. Hence the point-singular monopoles obey the quantization condition
\begin{equation}
\label{point} e\vec g = 2\pi n^i\,  \vec\alpha_i^\vee\quad  \quad  n^i=\hbox {integer}\, .
\end{equation} 
For  an $SO(3)$ theory with structure
constants $\epsilon^{abc}$, one recovers from Eq.(\ref{point})  the 't Hooft quantization condition for
the single component $g$ of the magnetic charge
\begin{equation}
\label{so3} e g = 4\pi n^i\quad  \quad  n^i=\hbox {integer}\, .
\end{equation}
One may then in general search
for regular monopole solutions by taking non-constant  values for the scalar fields and hence
admitting non-abelian configurations of the gauge fields. 

The lattice
$\Lambda_R^\vee$ generated by the coroots is a root lattice. It is
isomorphic to the original root lattice for all simple groups except the $C_n$ and $B_n$ series which
are interchanged, as pointed out by Goddard, Olive and Nuyts \cite {gno}. The transformation
$\alpha
\to
\alpha^\vee$ is an involution and one has thus in addition to the previous duality relation
$\Lambda_W^*=\Lambda_R^\vee$ the corresponding relation 
$\Lambda_R^*=\Lambda_W^\vee$. Solutions of Eq.(\ref{unobservable}) which are {\em not}
solutions of Eq.(\ref{nostring}) characterize Dirac monopoles. Solution of  Eq.(\ref{unobservable})
are on the lattice dual to $\Lambda_R$, hence their magnetic charges are given by    
\begin{equation}
\label{nadirac} e\vec g_d = 2\pi n^i\,  \vec m_i^\vee\quad  \quad  n^i=\hbox {integer} \quad,\quad
e\vec g_d
\neq 2\pi n^i\, 
\vec\alpha_i^\vee\, .
\end{equation}

 If ${\cal H}=\cal T$, the magnetic charges Eq.(\ref {point}) and (\ref{nadirac}) are well defined  (up to
Weyl reflections) but if
${\cal H}$ is larger than
${\cal T}$, these solutions can be continuously deformed
 in ${\cal H}$ and only some components can be defined in a topologically invariant way. If the
symmetry is fully broken, topological stability is  lost.  However    possibly regular locally stable
flux tubes can be formed and retain from the Dirac quantization condition Eq.(\ref{nadirac})
 the quantum numbers characterizing the distinct discrete  conjugation classes or equivalently
 the center of the group.
 
The duality relations between $\Lambda_W$ and
$\Lambda_R^\vee$ (and $\Lambda_R$ and $\Lambda_W^\vee$) was interpreted in reference
\cite{gno}    as an electric-magnetic duality between different gauge groups and was  generalized  
to all groups $\cal G$ locally isomorphic to $\tilde {\cal G}$. A conjecture of how  eletromagnetic
duality could be realized  in a full quantized theory for the BPS limit of  regular monopoles \cite{bps}
was suggested by Montonen and Olive
\cite {mo} and a form of the Montonen-Olive duality was  displayed in  $N=2$ supersymmetric
Yang-Mills theory~\cite{sw}.

These results give credence to the old conjecture that confinement is essentially magnetic
superconductivity \cite{confine}.  The BEH mechanism, when $\cal G$ symmetry is completely 
broken,  is a relativistic analog of superconductivity and may be viewed as a condensation of
electric charges. Magnetic fluxes are then channeled into  quantized  flux tubes.  In confinement, it is
the electric flux which is channeled into quantized tubes.  Electric-magnetic dualities suggest that,
at some fundamental level,  confinement is  a condensation of magnetic monopoles and constitutes
the magnetic dual of the BEH mechanism.

\subsection{Fermions from bosons}

The monopole solution Eq.(\ref{regular}) and its regular generalization are invariant under
simultaneous rotations in space and isospace. This is an invariance under the diagonal
subalgebra
$so_{ diag}(3)={\rm diag}[so_{space}(3)\oplus so_{isospace}(3)]$. It implies that a bound state of a
scalar of isospin
$1/2$ with the monopole is a space-time fermion
\cite{fermions}. In this way, fermions can be made out of bosons.

In field theory, such transmutations are rather exceptional. But it may be of  importance if the
nature of space-time emerges from a more basic description as illustrated in the string theory
approach to quantum gravity.  A suggestion along these lines was first proposed in reference \cite
{freund}. To see how this might happen, compactify  the bosonic closed string on maximal toroids  of
the rank 16 group
$E_8\times
\widetilde{SO}(16)/Z_g$, where
$Z_g$ is a subgroup of the center
$Z_2\times Z_2$ of the universal covering $\widetilde{SO}(16)$ of the rotation group
$SO(16)$. This yields four modular invariant bosonic closed string theories \cite {eh}. Each sector of
these closed strings contains  ten dimensional {\em space-time} fermionic subspaces, which appear
by selecting in the light-cone gauge  transverse states transforming under the diagonal
subalgebra  $so_{ diag}(8)={\rm diag}[so_{space}(8)\oplus so_{internal}(8)]$. Here
$ so_{internal}(8)$ is a subalgebra of the algebra $so(16)$ which  emerges as a symmetry of the 
compactified bosonic string. Consistency of the truncation to these states  stems from the
non-trivial requirement that the  algebra $so_{ diag}(8)$ closes on the Lorentz algebra
$so(9,1)$~\cite{gen, eh}. For spinor  representation of $so_{internal}(8)$, one gets
space-time fermions in analogy with the transmutation arising from the diagonal subalgebra
${\rm diag}[so_{space}(3)\oplus so_{isospace}(3)]$  in the 't Hooft-Polyakov
monopole. 

One obtains in this way all the consistent fermionic ten-dimensional closed strings, namely the
supersymmetric IIA and IIB and the non supersymmetric OA and OB  strings. One gets,
solely from bosonic consideration, the spectra and  tensions of   all their
$p$-dimensional D$p$-branes  as well as all their anomaly-free open descendants with the concomitant
Chan-Paton factors. All the fermionic strings are interrelated at the level of their bosonic parents
through the global properties of the sixteen-dimensional rotation group \cite {eh}.

Although these results are essentially kinematical in character, they  raise  the possibility
that  space-time fermions and perhaps even supersymmetry could arise from bosonic degrees of
freedom.  In such a perspective no fermionic degrees of freedom might be needed in a fundamental
theory of quantum gravity.

\subsection{\bf  A geometrical view on the BEH mechanism}

The BEH mechanism operates within the context of gauge theories. Despite the fact that grand
unification schemes reach scales comparable to the Planck scale, there was, a priori, no indication
that Yang-Mills fields offer any insight into    quantum gravity. The  superstring and M-theory
approach to quantum gravity did produce   theoretical achievements, in particular in the context of a
quantum interpretation of the black holes entropies. Of particular interest  in that context are
the D$p$-branes.  Here I will recall how D$p$-branes yield a geometrical interpretation of the BEH
mechanism. 

When $N$  BPS D$p$-branes coincide, they admit  massless excitations  from the 
$N^2$ zero length oriented strings with both end attached on the $N$ coincident branes. There are
$N^2$ massless vectors and additional $N^2$ massless scalars for each dimension transverse to the
branes. The open string sector has local $U(N)$ invariance. At rest,  BPS D$p$-branes  can separate
from each other in the transverse dimensions   at no cost of energy. Clearly this can break the
symmetry group from $U(N)$ up to
$U(1)^N$ when all the branes are at distinct location in the transverse space, because strings joining
two different branes have finite length and hence now describe finite mass excitations.  The only
remaining massless excitations are then due to the zero length strings with both ends  on the same
brane.  

\vskip .5cm
\hskip 4cm\epsfbox{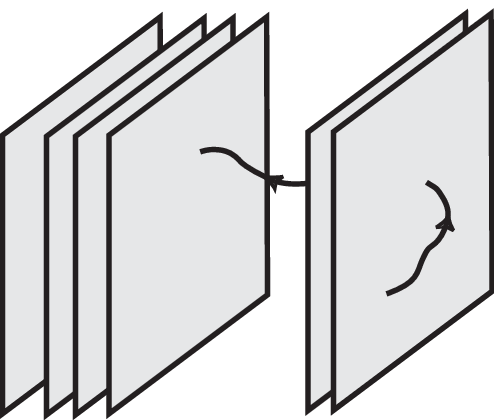}
\begin{quote}
\begin{center}
\baselineskip 10pt {\footnotesize Fig.9. Breaking $U(N)$ gauge symmetry by D$p$-branes.}
\end{center}
\end{quote}
\vskip .2cm

This symmetry breaking mechanism can be understood as a BEH mechanism from the  action
describing   low energy excitations of
$N$ D$p$-branes. The  action  is the  reduction to
$p+1$ dimensions of 10-dimensional supersymmetric Yang-Mills with
$U(N)$ gauge  fields \cite {reduc}. 

The Lagrangian is
\begin{equation}
\label{reduction} {\cal L}=- {1\over4}{\rm Tr}\,  {\bf F}_{\mu\nu} {\bf F}^{\mu\nu}+  {\rm Tr}
\left(~  {1\over2} D_\mu{\bf A}^iD^\mu{\bf A}^i - {1\over4} [{\bf A}^i\, , {\bf A}^j]^2  ~\right)+
\hbox{fermions} \, ,
\end{equation} where $\mu$ labels the $p+1$ brane coordinates and $i$ the directions transverse to
the branes. ${\bf F}_{\mu\nu}=F_{\mu\nu}^a {\bf T}_a $, ${\bf A}^i=A^{i\, a}~ {\bf T}_a$ where ${\bf
T}_a$ is a generator of $U(N)$ in a defining representation.

The states of zero energy are given classically, and hence in general because of supersymmetry, by
all commuting
${\bf A}^i =\{x^i_{mn}\}$  matrices, that is, up to an equivalence, by all diagonal matrices
$\{x^i_{mn}\}=\{x^i_{m}\delta_{mn}\}$. Label the $N^2$ matrix elements of ${\bf A}_\mu$ by
$A_{\mu\, mn}$. The 
$(N^2-N)$ gauge  fields given by the non diagonal elements $m\neq n$ acquire a mass
\begin{equation}
\label{nd} m^2_{mn}\propto (\vec x_m -\vec x_n)^2 \, ,
\end{equation} if $\vec x_m\neq\vec x_n $, as   is easily checked by computing  the quadratic  terms
in $A_{\mu\, mn}$ appearing in the covariant derivatives ${\rm Tr} D_\mu{\bf A}^iD^\mu{\bf A}^i$. 

This symmetry breaking is
 induced by  the  expectation values 
$\{x^i_{m}\}$. The gauge invariance is ensured, as usual, by unobservable ($N^2-N$) would-be NG
bosons. To identify the latter,  consider the scalar potential in Eq.(\ref{reduction}), namely
\begin{equation}
\label{scalar} V= {\rm Tr} {1\over4} [{\bf A}^i\, , {\bf A}^j] [{\bf A}^i\, , {\bf A}^j] =
{1\over4}\sum_{i,j;m,n}
\langle m\vert [{\bf A}^i\, , {\bf A}^j] \vert n \rangle \langle n\vert [{\bf A}^i\, , {\bf A}^j]\vert
m\rangle\, .
\end{equation} One writes
\begin{equation}
\langle m\vert {\bf A}^j \vert n \rangle = x^j_{m}\delta_{mn} + y^j_{mn}\, .
\end{equation} Here the   diagonal elements
$\{x^j_{m}\}$ are the  expectation values and the  $y^j_{mn}(= - [y^j_{nm}]^\star)$ define
$d(N^2-N)$ hermitian scalar fields  $(y^i_{mn})^a ~ (a= 1,2)$ where $y^j_{mn} = (y^j_{mn})^1 + i
(y^j_{mn})^2 ~, ~ m>n$~,  and
$d$ is the number of transverse space dimensions. The mass matrix for the fields
$(y^i_{mn})^a $ is
\begin{equation}
\label{matrix} {\partial ^2 V\over \partial (y^k_{mn})^a \partial (y^l_{mn})^b }= \delta^{ab}[(\vec
x_m -\vec x_n)^2
\delta^{kl}- (x^k_{m}-x^k_{n})(x^l_{m}-x^l_{n})]\, ,
\end{equation} and has for each  pair $m,n ~(m<n)$, two zero eigenvalues corresponding to the
eigenvectors  
$ (y^l_{mn})^a\propto (x^l_{m}-x^l_{n})$. These are the required $(N^2-N)$ would-be NG bosons, as
can be checked directly from the coupling of ${\bf A}^i$ to  ${\bf A}_\mu$ in the Lagrangian
Eq.(\ref{reduction})~. 

As mentioned above, the breaking of $U(N)$ up to
$U(1)^N$  may be viewed in the string picture
 as due to  the stretched  strings joining branes separated in the
 dimensions transverse to the branes. One identifies the $\{x^i_{m}\}$ as 
 coordinates transverse to the brane
$m$. The mass of  the vector  $A_{\mu\, mn}$ is then the mass shift, due to the  stretching,  of
the otherwise massless  open string vector excitations.   The unobservable  NG bosons $\vec
y_{mn}\parallel (\vec x_m -\vec x_n)
$ are  the field theoretic expression of the unobservable longitudinal modes of the strings joining
the branes $m$ and $n$.   In this  way D$p$-branes provide a geometrical interpretation of the BEH
mechanism.

An  interesting situation  occurs when
$p=0$ \cite {reduc, bfss}. The Lagrangian Eq.(\ref{reduction}) then describes
 a pure quantum mechanical system where the $\{x^i_{mn}\}$ are the dynamical
variable\footnote{This Lagrangian  first appeared as a description of the supermembrane
\cite{super}.}. The time component 
${\bf A}_t$ which enters the covariant derivative $D_t {\bf A}^i$ can be put equal to zero, leaving a
constraint which amounts to restrict the quantum states to singlets of $SU(N)$. The $\{x^i_{m}\}$
which define in string theory D$0$-brane coordinates (viewed as partons in the infinite momentum
frame in reference~\cite{bfss}) are  the analog, for $p=0$, of  the  expectation values in the
$p\neq0$ case, although  they label now classical collective position variables of the quantum
mechanical system. The non-diagonal quantum degrees of
freedom 
$\vec y_{mn}~\bot~ (\vec x_m -\vec x_n)$  have  a positive potential  energy proportional to the
distance squared between the D$0$-branes $m$ and
$n$.  Hence they get locked in their ground state when the D$0$-branes are
 largely separated  from each other.  In this way, the D0-brane
${\bf A}^i =\{x^i_{mn}\}$  matrices  commute at large distance scale and define geometrical degrees
of freedom. However these  matrices  do not commute at short distances where the potential
energies of the
$y^i_{mn}$ go to zero. This, and its aforementioned analog for D$p$-branes ($p>0$) suggests that the
space-time geometry exhibits non commutativity at small distances \cite{reduc}, a feature which
might  be relevant for  quantum gravity.

$$--------- $$

In the second half of the twentieth century, the progress of our understanding of natural phenomena 
in rational terms bears the mark of Yang-Mills local gauge invariance. The reconciliation of this large
symmetry  with the apparent diversity of natural phenomena where symmetry is  hidden appears
possible through the implementation  of a structured vacuum originating in the concept of
spontaneous broken symmetry.

\section*{\bf  Acknowledgments}

Robert Brout was my teacher and became my friend. In writing these notes I revive the
 intensity
 of our early collaborations when I caught the first glimpse of his unique way of perceiving logical
principles in physical terms.


\begin{thebibliography}{99}

\bibitem{landau} L.D. Landau,  {\it On the theory of phase transitions I},  Phys. Z. Sowjet. {\bf 11}
(1937) 26  [JETP {\bf 7} (1937) 19].

\bibitem{nambu2} Y. Nambu, {\it Axial vector current conservation in weak interactions}, Phys. Rev.
Lett. {\bf 4} (1960) 380.

\bibitem{nambujl} Y. Nambu and G. Jona-Lasinio, {\it Dynamical model of elementary particles based
on an analogy with superconductivity I, II}, Phys. Rev. {\bf 122} (1961) 345; Phys. Rev. {\bf 124}
(1961)  246.

\bibitem{eb} F. Englert and R. Brout, {\it Broken symmetry and the mass of gauge vector mesons},
Phys. Rev. Lett. {\bf 13}  (1964) 321.

\bibitem{higgs2} P.W. Higgs, {\it Broken symmetries and the masses of gauge bosons}, Phys. Rev.
Lett. {\bf 13} (1964) 508. 

\bibitem{renorm} G. 't Hooft,  {\it Renormalizable Lagrangians for massive Yang-Mills fields}, Nucl.
Phys. {\bf B35} (1971) 167;   G. 't Hooft and M. Veltman,  {\it Regularization and renormalization of
gauge fields}, Nucl. Phys. {\bf B44} (1972) 189.

\bibitem{gws} S.L. Glashow, {\it Partial-symmetries of weak interactions}, Nucl. Phys. {\bf 22} (1961)
579;  S. Weinberg, {\it A model of leptons}, Phys. Rev. Lett.  {\bf 19} (1967) 1264; A. Salam, 
Proceedings of the 8th Nobel Symposium, {\it  Elementary  Particle Physics},  ed. by N.
Svartholm,(Almqvist and Wiksell, Stockhlom) p 367.

\bibitem{bcs} J. Bardeen, L. Cooper and J.R. Schrieffer, {\it Microscopic theory of superconductivity},
Phys. Rev. {\bf 106} (1957) 162.

\bibitem{anderson}P.W. Anderson, {\it Random-phase approximation in the theory of
superconductivity}, Phys. Rev. {\bf 112} (1958) 1900.

\bibitem{nambu1}Y. Nambu, {\it Quasi-particles and gauge invariance in the theory of
superconductivity}, Phys. Rev. {\bf 117} (1960) 648.

\bibitem{goldstone} J. Goldstone, {\it Field theories with ``superconductor'' solutions}, Il Nuovo
Cimento {\bf 19} (1961) 154.

\bibitem{gsw}  J. Goldstone, A. Salam and S. Weinberg, {\it Broken symmetries}, Phys. Rev.  {\bf 127}
(1962) 965.

\bibitem{elitzur} S. Elitzur, {\it Impossibility of spontaneously breaking local symmetries}, Phys.
Rev. {\bf D12} (1975) 3978.

\bibitem{higgs1} P.W. Higgs,  {\it Broken symmetries, massless particles and gauge
fields}, Phys. Lett. {\bf 12} (1964) 132.

\bibitem{higgs3} P.W. Higgs, {\it Spontaneous symmetry breakdown without massless
bosons},  Phys. Rev. {\bf 145} (1966) 1156.

\bibitem{be2} F. Englert, R.Brout and M. Thiry, {\it Vector mesons in presence of broken symmetry}, Il
Nuovo Cimento {\bf 43A} (1966) 244;  Proceedings of the 1997 Solvay Conference, {\it Fundamental
Problems in Elementary  Particle Physics},  Interscience Publishers J. Wiley and Sons, p 18.

\bibitem{gqw} H. Georgi, H.R. Quinn and S. Weinberg, {\it Hierarchy of interactions in unified gauge
theories}, Phys. Rev. Lett. {\bf 33} (1974) 451.

\bibitem{dirac} P.A.M. Dirac, {\it The theory of magnetic poles}, Phys. Rev. {\bf 74} (1948) 817.

\bibitem{hp} G 't Hooft, {\it Magnetic monopoles in unified gauge theories}, Nucl. Phys. {\bf B79}
(1974) 276; A.M. Polyakov, {\it Particle spectrum in the quantum field theory}, Pisma
Zh. Eksp. Teor. Fiz. {\bf 20} (1974) 430  [JETP Lett. {\bf 20} (1974) 194].

\bibitem{all}  F. Englert and P. Windey, {\it Quantization condition for 't Hooft monopoles in compact
simple Lie groups}, Phys. Rev. {\bf D14} (1976) 2728.

\bibitem{wy} T.T. Wu and C.N. Yang,  {\it Concept of nonintegrable phase factors and global
formulation of gauge fields}, Phys. Rev. {\bf D12} (1975) 3845.

\bibitem{gno} P. Goddard, J. Nuyts and D. Olive, {\it Gauge theories and magnetic charges}, Nucl. Phys.
{\bf B125} (1977) 1.

\bibitem{bps} M.K. Prasad and C.M. Sommerfield, {\it Exact classical solution for the 't Hooft
monopole and the Julia-Zee dyon},  Phys. Rev. Lett. {\bf 35} (1975) 760; E.B. Bogomolny, {\it Stability
of classical solutions}, Sov. J. Nucl. Phys. {\bf 24} (1976) 449. 

\bibitem{mo} C. Montonen  and D. Olive, {\it Magnetic monopoles as gauge particles?}, Phys.Lett. {\bf
B72} (1977)  117.

\bibitem{sw} N. Seiberg and E. Witten, {\it Electric-magnetic duality, monopole condensation, and
confinement in N=2 supersymmetric Yang-Mills theory}, Nucl.Phys. {\bf B426} (1994) 19;  Erratum,
{\bf B430} (1994) 485, {\tt arXiv: hep-th/9407087}. 

\bibitem {confine} S. Mandelstam, {\it Vortices and quark confinement in non-Abelian gauge
theories}, Phys. Rep. {\bf 23C} (1976) 245; F. Englert and P. Windey, {\it Electric confinement and
magnetic superconductors}, Nucl.Phys. {\bf B135} (1978) 529;  G 't Hooft, {\it On the phase
transition towards permanent quark confinement}, Nucl. Phys. {\bf B138} (1978) 1.

\bibitem{fermions} R. Jackiw and C. Rebbi, {\it Spin from isospin in a gauge theory},
Phys. Rev. Lett. {\bf 36} (1976) 1116; P. Hasenfratz and G. 't Hooft,   {\it Fermion-boson puzzle in a
gauge theory}, Phys. Rev. Lett. {\bf 36} (1976) 1119; A. Goldhaber,  {\it Connection of spin and
statistics for charge-monopole composites}, Phys. Rev. Lett. {\bf 36} (1976) 1122.

\bibitem{freund} P.G.O. Freund,  {\it Superstrings from 26 dimensions?}, Phys.Lett. {\bf B151} (1985)
387.

\bibitem{eh} A. Chattaraputi, F. Englert, L. Houart and A. Taormina, {\it The bosonic mother of
fermionic D-branes},
  JHEP {\bf 0209} (2002) 037, {\tt arXiv:hep-th/0207238}. 

\bibitem{gen} A. Casher, F. Englert, H. Nicolai and A. Taormina, {\it Consistent superstrings as
solutions of the D= 26 bosonic string theory}, Phys. Lett. {\bf B162} (1985) 121;  F. Englert, H. Nicolai
and A.N. Schellekens, {\it Superstrings from 26 dimensions}, Nucl. Phys. {\bf 274}  (1986) 315;  W.
Lerche, D. L\" ust and A. Schellekens, {\it Ten-dimensional heterotic strings from
Niemeier lattices}, Phys. Lett.  (1986) {\bf B181} 71; Erratum,  Phys. Lett. {\bf B184} (1987) 419.

\bibitem{reduc}  E. Witten, {\it Bound states of strings and p-branes}, Nucl. Phys. {\bf B460} (1996)
335,
 {\tt arXiv:hep-th/9510135}.

\bibitem{bfss} T. Banks, W. Fischler, S.H. Shenker and L. Susskind, {\it M theory as a matrix model: A
conjecture}, Phys.Rev, {\bf D55} (1997)
5112, {\tt arXiv:hep-th/9610043}.

\bibitem{super} B. de Wit, J. Hoppe and H. Nicolai, {\it On the quantum mechanics of
supermembranes}, Nucl. Phys. {\bf B305} (1988) 545.

\end{thebibliography}
\end{document}